\documentclass[amsart,11pt]{article}
\usepackage{cite}
\usepackage{amsmath,amssymb,amsfonts}
\usepackage{algorithmic}
\usepackage{graphicx}
\usepackage{textcomp}
\usepackage{verbatim}
\usepackage{xcolor}
\usepackage{geometry}
\usepackage{authblk}

\newcommand{\R}{\mathbb R}
\newcommand{\F}{\mathcal{F}}
\newcommand{\T}{^\mathsf{T}}

\newcommand{\mL}{\mathcal L}

\newcommand{\x}{\vec{x}}

\def\dd{\mathrm{d}}

\def\vvec#1{\vec{\vec{#1}}}

\usepackage[normalem]{ulem}

\title{A Field Free Line 3D Reconstruction Model for Magnetic Particle Imaging for Improved Sensitivity, Resolution, and High Dynamic Range Imaging
}

\author[1]{Toby Sanders\thanks{Corresponding author: sandertl20@gmail.com}}
\author[2]{Hayden Carlton}
\author[2]{Preethi Korangath}
\author[1]{Olivia C. Sehl}
\author[2]{Robert Ivkov}
\author[1]{Patrick W. Goodwill\thanks{Research reported in this publication was supported by the National Institute of Biomedical Imaging and Bioengineering and the National Cancer Institute of the National Institutes of Health under award numbers 1R44EB035078, R44CA285064, R44EB029877, R01CA257557, and S10OD026740. The authors would also like to thank the Jayne Koskinas and Ted Giovanis Foundation for Health and Policy for the support on this project. The content is solely the responsibility of the authors and does not necessarily represent the official views of the National Institutes of Health.}}

\affil[1]{Magnetic Insight, Alameda, CA, USA}
\affil[2]{Johns Hopkins School of Medicine, Baltimore, MD, USA}

\date{}

\begin{document}

\maketitle

\begin{abstract}
Magnetic particle imaging (MPI) is a tracer-based imaging modality that detects superparamagnetic iron oxide nanoparticles in vivo, with applications in cancer cell tracking, lymph node mapping, and cell therapy monitoring. We introduce a new 3D image reconstruction framework for MPI data acquired using multi-angle field-free line (FFL) scans, demonstrating improvements in spatial resolution, quantitative accuracy, and high dynamic range performance over conventional sequential reconstruction pipelines. The framework is built by combining a physics-based FFL signal model with tomographic projection operators to form an efficient 3D forward operator, enabling the full dataset to be reconstructed jointly rather than as a series of independent 2D projections. A harmonic-domain compression step is incorporated naturally within this operator formulation, reducing memory overhead by over two orders of magnitude while preserving the structure and fidelity of the model, enabling volumetric reconstructions on standard desktop GPU hardware in only minutes. Phantom and in vivo results demonstrate substantially reduced background haze and improved visualization of low-intensity regions adjacent to bright structures, with an estimated $\sim$11$\times$ improvement in iron detection sensitivity relative to the conventional X-space CT approach. These advances enhance MPI image quality and quantitative reliability, supporting broader use of MPI in preclinical and future clinical imaging.
\end{abstract}

\section{Introduction}
Magnetic particle imaging (MPI) is an emerging imaging modality that directly detects iron oxide magnetic nanoparticles (MNPs), offering strong potential for a variety of biomedical applications including cancer cell tracking\cite{zheng2015magnetic,sehl2020perspective}, lymph node mapping\cite{sehl2024first}, and cell therapy monitoring\cite{weizenecker2009three, panagiotopoulos2015magnetic,talebloo2020magnetic}. On commercial scanners, 3D image reconstruction is commonly performed by first processing 2D projection data individually and then combining the resulting images into a 3D volume using computed tomography (CT) methods\cite{konkle2012projection}. However, sequential processing of 2D projections can limit image quality. In particular, information unique to each scan—such as system nonidealities and temporal fluctuations—is not fully leveraged when projections are reconstructed independently. These nonidealities can arise from a range of sources: temperature-induced amplifier drift, mechanical vibrations altering receive coil geometry, and variable tissue loading affecting coil sensitivity patterns. A reconstruction framework that simultaneously incorporates all 3D data could better account for these variations, potentially leading to improved image fidelity, sensitivity, and signal-to-noise ratio.

\begin{figure}
	\centering
	\includegraphics[width=1\textwidth]{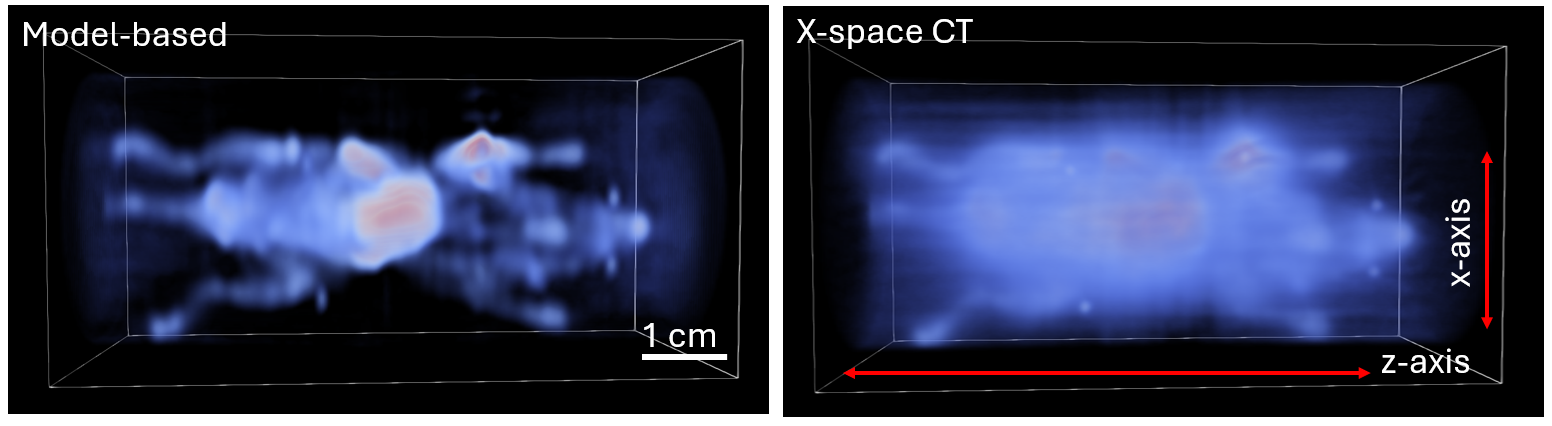}
	\caption{3D volume renderings of in vivo MPI reconstructions comparing the proposed model-based 3D FFL reconstruction (left) to the conventional sequential X-space CT reconstruction (right). A cool-to-warm colormap is used, with low-intensity values shown in light blue and high-intensity values shown in red. The new model-based reconstruction exhibits improved spatial resolution and reduced background haze, particularly around high-signal regions such as the liver and tumor. This enhancement enables more accurate quantification of nearby low-intensity regions that are often obscured in the conventional reconstruction.\\}
	\label{fig: mouse3D}
\end{figure}

Despite this potential, full 3D reconstruction directly from time-domain MPI data has remained largely impractical for large fields of view due to memory and computational constraints. Scaling from 2D to 3D increases the data size dramatically; uncompressed time-domain data from a single 3D scan can exceed 100 GB, placing a substantial burden on modern hardware.

To overcome these challenges, we present a novel solution comprising two key innovations. First, we introduce a 3D field-free line (FFL) reconstruction model that integrates multiple 2D projections into a coherent 3D imaging framework using efficient tomographic and physics-based operators. Second, we develop a time-domain data compression method tailored to MPI signals, which are naturally concentrated in narrow frequency bands. This compression approach significantly reduces memory requirements while retaining essential signal content, enabling efficient computation on standard GPUs.

Figure \ref{fig: mouse3D} provides a motivating example: 3D reconstructions of an in vivo mouse using both the conventional X-space CT method and our new model-based approach. The new method clearly improves spatial resolution and reduces background haze, particularly in regions of high signal intensity like the liver and tumor, thereby enhancing visibility of adjacent low-intensity structures.

In the remainder of this article, we derive the reconstruction model, describe its implementation, benchmark its computational performance, and evaluate its image quality on both phantom and in vivo data. Our results demonstrate that this method achieves significant improvements in resolution, sensitivity, and dynamic range. This is particularly highlighted in challenging scenarios such as imaging small metastases near high-signal regions, marking a substantial advance in MPI reconstruction methodology.

\subsection{Previous Related Work}
Image reconstruction in MPI has been the focus of significant research over the past decade \cite{zhang2024current}. Two of the leading approaches are X-space methods\cite{goodwill2010x, goodwill2011multidimensional} and system matrix (SM) calibration\cite{knopp2015local,szwargulski2018efficient}. While SM calibration can model the imaging process with high fidelity, it requires extensive calibration using point source phantoms for each new scan sequence or MNP type. This makes it impractical in many settings and typically results in coarse, pixelated reconstructions due to limitations in calibration resolution.

X-space methods offer a direct reconstruction framework by stitching together the scan sequence’s time-domain data into the imaging domain. However, these methods are limited in utility as they are not inherently model-based and are highly sensitive to noise. Although simulation studies have proposed generalizations of X-space for more flexible scan geometries \cite{ozaslan2019fully}, such extensions remain unvalidated experimentally.

Various model-based reconstruction techniques have also been proposed. These include reconstructions using Chebyshev polynomial bases\cite{droigk2022direct} and forward models built from detailed MPI system simulations\cite{knopp2009model,knopp20102d}. Such methods have been further refined to include complex physical behaviors like magnetic relaxation\cite{kluth2018mathematical,maass2024equilibrium,weizenecker2018fokker}. While these enhancements offer interesting theoretical insights, their practical impact on image quality remains uncertain. In addition, accurately implementing these more detailed models in clinical environments poses challenges, as MNP behavior can vary with manufacturing batch, anatomical location, and time post-injection\cite{salimi2025mpi,guzy2020complex,imhoff2025characterization}.

Our approach builds on the standard Langevin paramagnetic model, which omits relaxation and interparticle interactions, reducing the number of MNP-specific parameters to a single magnetic moment and its associated saturation parameter. This simpler modeling framework is aligned with earlier system simulation-based models\cite{knopp2009model,knopp20102d} but is restructured here into a highly efficient operator-based format. Each operator in the model is computationally tractable and memory-efficient. As a result, our method can readily accommodate changes in scan parameters, drive amplitudes, and MNP types, while enabling high-resolution reconstructions across 2D and 3D imaging meshes.

We believe this method establishes a new standard for computationally efficient, high-resolution MPI reconstruction. It provides a flexible foundation that could be extended with more detailed MNP physics in the future, although current simulations already show excellent agreement with experimental results\cite{sanders2025physics}.
This work extends that prior framework to handle the more complex 3D FFL reconstruction problem and demonstrates clear improvements in image quality over conventional methods.

\subsection{Background Notation and Theory}
We begin the technical discussion by providing some basic background notation and theory behind an MPI scanner. A comprehensive summary of the notation used is also provided in Appendix \ref{sec: notation} For a more detailed characterization of MPI, see references\cite{goodwill2011multidimensional,knopp2017magnetic,knopp2012magnetic}.

The time varying magnetic field generated for MPI is given by 
\begin{equation}\label{eq: H0}
	H(\vec x , t) = H_D(t) - G\vec x ,	
\end{equation}
where $H_D(t)$ is excitation drive field, $G$ is the gradient matrix, $\vec x = (x,y,z)\T$ is the spatial position variable, and $-G\vec x$ is the static magnetic field. For an FFL geometry with the FFL along the $y$-axis\cite{goodwill2012projection}, the static gradient matrix is given by 
\begin{equation}\label{eq: G}
	G = \begin{bmatrix}
		-G_0 & 0 & 0\\ 0 & 0 & 0 \\ 0 & 0 & G_0
	\end{bmatrix}.
\end{equation}
The drive field $H_D$ has no component along the $y$-axis, so that the FFL will traverse in the $xz$ planes over time. The drive field on the scanner is an oscillating sinusoidal wave, either in $z$ or $x$. In $z$, this is given by
\begin{equation}\label{eq: HD}
	H_D(t) = (0, 0, A \sin(2\pi f_0 t)),
\end{equation}
where $A$ is proportional to the drive amplitude and $f_0$ is the fundamental drive frequency.

The FFL at time $t$ is defined by the set of spatial locations where $H(\vec x , t) = 0$, which we denote by $\vec \xi(t)$. Then the total magnetic field can be expressed as
\begin{equation}\label{eq: H1}
	H(\vec x , t) = G (\vec \xi(t) - \vec x),
\end{equation}
which can be realized by plugging $\vec \xi (t)$ into (\ref{eq: H0}) and setting to zero.

Writing the MNP density map as $\rho(\x)$, the MNPs {average} response to the applied magnetic field is given by \cite{ferro1985}
\begin{equation}
	M(\x , t ;H) = m \rho(\x ) {\mL \left[\beta \| H (\x , t ) \| \right] } {\frac{H(\x , t)}{\| H(\x , t) \|} },
\end{equation}
where $\mL$ is the Langevin function, $m[Am^2]$ is the magnetic moment of a single MNP, and $\beta$ is a conglomerate constant related to the system parameters and MNP properties \cite{goodwill2011multidimensional}. The receive coil detects the total change of this field over time, with respect to spatial sensitivities inherent to the receive coil defined as $\vec b(\x) = (b_x (\x ) , b_y (\x ) , b_z (\x ) )\T $. The unfiltered received signal for this coil is given by
\begin{equation}\label{eq: s01}
	\begin{split}
		s_0(t)
		& =\frac{d}{dt} \iiint \vec b(\vec x)\T M(\vec x , t; H)\, \dd x \, \dd y\, \dd z
	\end{split}
\end{equation}

As shown in \cite{goodwill2011multidimensional}, the time derivative in (\ref{eq: s01}) can be analytically evaluated to obtain
\begin{equation}\label{eq: reals0}
	\begin{split}
		s_0(t) 
		& = m \iiint \rho(\x) \vec{b}\T(x) \vvec{h} (\vec \xi (t) - \x)   \vec v(t) \, \dd \x \\
		&  = m \sum_{i,j=1}^3 \iiint \rho(\x ) v_j (t) b_i (\x ) h_{ij} (\vec \xi(t) - \x ) \, \dd \x,
	\end{split} 
\end{equation}
where $\vec v(t) = \vec \xi'(t)$ is the velocity of the FFP and $\vvec{h}$ is a $3\times3$ tensor function (see Appendix \ref{sec: notation} for the definition). We typically refer to $\vvec{h}$ as the MPI point spread functions (PSFs), and additional details can be found references\cite{goodwill2011multidimensional, sanders2025physics}.  

	Finally, the noise-free received signal is modeled by
	\begin{equation}
		s(t) = s_0 * \gamma(t) ,
	\end{equation}
	where $\gamma(t)$ is a time domain notch filter that filters out information at the drive field frequency.

	\subsection{Review of MPI Physics-based Computational Model}\label{sec: derive}
	A discretized physics-based computational imaging model for MPI derived in our previous work\cite{sanders2025physics} is given by
	\begin{equation} \label{eq: svecHPF}
		\vec s =   \mathbf{\Gamma VEH B}  \rho,
	\end{equation}
	where each of the bold uppercase letters in this equation represents a matrix-based operator relating to the individual component of the received signal model. The complete forward model operator we simply write as $\mathbf{M = \Gamma VEH B}$. A summary of all of the operators in the model are given in Table \ref{table: opers}, which also provides the relationship between the operators and the functions involved. The decomposition of the MPI signal model into these particular components was carefully designed so that each of the operators can be evaluated efficiently, i.e. very fast and without major memory requirements. This model is valid for both 2D FFL projection imaging and 3D FFP volumetric imaging.

	\begin{table}[ht]
		\centering
		\begin{tabular}{| c | c | c | }
			\hline 
			Operator 				   &	Function Form & Description   \\ \hline
			$\mathbf B$  			 & $\vec b (\vec x)$  & Receive field sensitivity \\
			$\mathbf H$  			 & $\{ h_{ij} (\vec x) \}_{i,j}$ & Langevin PSFs  \\
			$\mathbf E$  			 & $\vec \xi(t)$     & FFP selection \\
			$\mathbf V$  			 &  $\vec v(t) $       & Velocity multiplier  \\
			$\mathbf \Gamma $  & $\gamma (t)$  & Receive chain filter  \\
			$ \mathbf M = \mathbf{\Gamma VEHB} $ & eq. (\ref{eq: reals0}) & Complete MPI model  \\ \hline
		\end{tabular}
		\caption{\label{table: opers}Summary of all of the efficient operators that are used to build the MPI model and solve the associated inverse problem.}
	\end{table}

	\section{Derivation of the New 3D FFL Model}
	\subsection{Harmonic Compression of the Model}\label{sec: compress}
\begin{figure}
		\centering
		\includegraphics[width=0.85\textwidth]{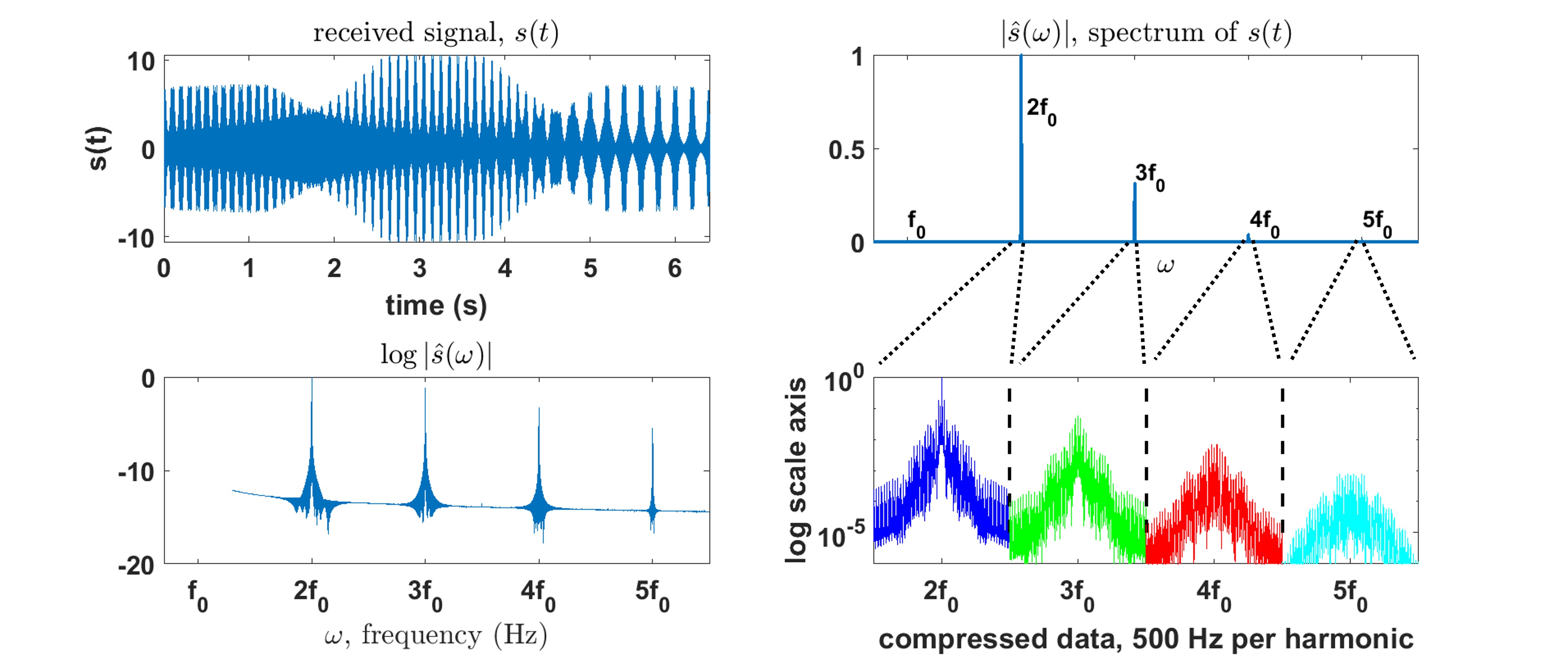}
		\caption{Compression of the time domain signal. Top left: simulated time domain MPI signal data. Top right: spectrum magnitude of the MPI signal data. Bottom left: logarithm of the spectrum magnitude. Bottom right: final compressed data form used for the reconstruction. The compressed form of the data reduces to total data memory overhead by 99.90\%.}
		\label{fig: comp}
	\end{figure}

	In this section we show how to compress the existing time domain model in (\ref{eq: svecHPF}) with only a few operations. This is possible because the energy of an MPI signal is highly concentrated around the signal's harmonics (see Figure \ref{fig: comp} and references\cite{sanders2025multi} for details), which are the Fourier domain frequency bands around the integer multiples of the fundamental drive field frequency as defined by (\ref{eq: HD}). This implies that the time domain signal data can be compressed for storage and processing purposes without significant loss of information. 

	For some harmonic integer $k\ge 2$, we can write the band of Fourier coefficient data around this harmonic as
	\begin{equation}
		s_k = \mathcal{S}_k \F   \vec s = \mathcal{S}_k \F \mathbf{M}\rho ,
	\end{equation}
	where $\F$ is the Fourier transform, and $\mathcal S_k$ is a selection operator containing a subset of the rows of the identity matrix, which selects the band of Fourier coefficients centered at the $k$th harmonic. Then we define the digitally compressed (DC) data and forward model operator by
	\begin{equation}\label{eq: bcompress}
		b_{DC} := \begin{bmatrix}
			s_2 \\ s_3 \\ \vdots \\ s_K
		\end{bmatrix} 
		,
		\,
		\mathcal S := 
		\begin{bmatrix}
			\mathcal{S}_2 \\ \mathcal{S}_3 \\ \vdots \\ \mathcal{S}_K
		\end{bmatrix},
		\,
		\text{and}
		\,\,
		\mathbf A_{DC} := \mathcal S {\F} \mathbf M,
	\end{equation}
	and so our DC model may be written as 
	\begin{equation}\label{eq: modelDC1}
		\begin{split}
			b_{DC} & = \mathbf A_{DC} \rho\\
			& = \mathcal S \mathcal F \mathbf{\Gamma V E H B}\rho
		\end{split}
	\end{equation} 
	The bottom right plot of Figure \ref{fig: comp} shows the amplitude of the final compressed data form, $b_{DC}$, for $K=5$.

	Finally, we may simplify this final operator one step further by considering the form of the receive chain filter, $\mathbf \Gamma$. This operation acts as an analog convolutional filter, which we can write as a Fourier domain multiplication by $\mathbf{\Gamma} = \F^{-1} D_\Gamma \F$, where $D_\Gamma$ is a diagonal matrix taking on the values (magnitude and phase) of the receive transfer function. Substituting this expression into (\ref{eq: modelDC1}), our DC model operator reduces to
	\begin{equation} \label{eq: DCmodel}
		\mathbf A_{DC} = 
		\mathcal S D_\Gamma \F
		\mathbf{ V E H B}.
	\end{equation}
	Although this model is mathematically equivalent to that in (\ref{eq: modelDC1}), it emphasizes that we only need to perform one FFT, as opposed to 3 FFTs required by naively implementing (\ref{eq: modelDC1}). As a consequence, this actually reduces the computational complexity of the original model-based operator. While this complete model may subjectively appear complex, it is intentionally written in a way that emphasizes computational components that are each efficient to compute. The reader may refer to previous works for further details on the computational methods used\cite{sanders2025physics, sanders2020effective}. 
	
	In our implementations, we have chosen to keep 0.5 kHz of bandwidth around each harmonic and harmonics 2 through 5, as shown in Figure \ref{fig: comp}. These choices were based in both empirical and statistical observations, as outlined by the compression analysis in Appendix \ref{sec: companaly} This was found to reduce the overall memory overhead of the data by 99.90\%.

	\subsection{3D FFL Model}\label{sec: 3DFFL}
\begin{figure}
		\centering
		\includegraphics[width=0.45\textwidth]{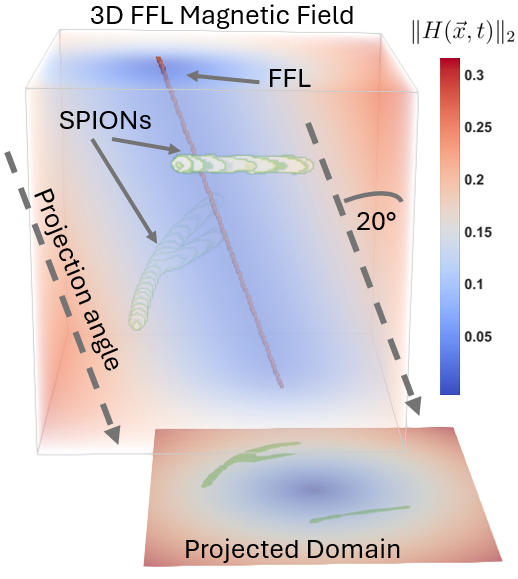}
		\caption{Diagram of the FFL geometry in 3D. At lines along the 20$^{\circ}$ FFL angle, the magnetic field is a constant. Therefore, when the transmit amplitude is orthogonal to this line, the full 3D MPI model can be constructed with a 3D to 2D projection operation followed by the 2D MPI model.}
		\label{fig: 3Ddiagram}
	\end{figure}

	The model re-introduced so far in this manuscript is valid for 2D projection imaging of FFL scanner data and 3D volumetric imaging of FFP scanner data. In this section, we show how to extend the 2D FFL projection imaging model at multiple angles to a full 3D volumetric imaging model. The primary underlying concept is to combine a parallel beam tomographic projection operator with the 2D FFL projection operator (see Figure \ref{fig: 3Ddiagram}). 
	
	For the FFL geometry as characterized in equations (\ref{eq: G}) and (\ref{eq: H1}), we can rewrite the signal in (\ref{eq: s01}) in the projection form as
	\begin{equation}
		\begin{split}
			s_0(t) = \frac{d}{dt} \iint & \left[ \int \rho(\x ) \, \dd y\right]  \vec b(\vec x')\T 
			{\mL \left[\beta \| H (\vec x' , t ) \| \right] } {\frac{H(\vec x' , t)}{\| H(\vec x' , t) \|} } \, \dd x \,  \dd z ,
		\end{split}
	\end{equation}
	where $\vec x' = (x,z)$ condenses $\vec x$ to only its $x$ and $z$ components. The model written above is effectively the 2D projection model used for MPI, i.e. to reconstruct the projected object within the brackets given by $\rho_0(\vec x') = \int \rho(\vec x) \, \dd y$. To collect 3D data for tomographic imaging, the magnetic field may be rotated around the $z$-axis\footnote{Rotation specifically around the $z$-axis is not required, but this is the model we construct to match our scanner configuration and to keep the details concise.} by angle $\theta$, and a new scan is acquired at this angle. It is mathematically equivalent to modeling the rotation of the MNP density, $\rho(\vec x)$, by $\theta$ and leave the magnetic model stationary. This gives rise to the following angular projection data:

	\begin{equation}\label{eq: ptheta}
		\begin{split}
			s_0(t ; \theta ) 
			& = \frac{d}{dt} \iint\left[ \int \rho((x,y)Q_\theta\T , z ) \, \dd y\right]   \vec b(\vec x')\T 
			{\mL \left[\beta \| H (\vec x' , t ) \| \right] } {\frac{H(\vec x' , t)}{\| H(\vec x' , t) \|} } \, \dd x \,  \dd z \\
			& = \frac{d}{dt} \iint\rho_\theta(\vec x')  \vec  b(\vec x')\T {\mL \left[\beta \| H (\vec x' , t ) \| \right] } {\frac{H(\vec x' , t)}{\| H(\vec x' , t) \|} } \, \dd x \,  \dd z
		\end{split}
	\end{equation}
	where 
	\begin{equation}\label{eq: qtheta}
		Q_\theta = \begin{bmatrix}
			\cos(\theta) & \sin(\theta) \\ - \sin(\theta) & \cos(\theta)
		\end{bmatrix},
	\end{equation}
	and 
	\begin{equation}\label{eq: ptheta2}
		\rho_\theta(\vec x') = \int \rho((x,y) Q_\theta\T , z) \dd y.
	\end{equation}
	The outer integrals in (\ref{eq: ptheta}) are modeled in the discretized domain by the same operators provided in \cite{sanders2025physics}, which are summarized in Section \ref{sec: derive}. We can therefore write the forward model mapping $\rho_\theta$ to $s_0(t;\theta)$ as
	\begin{equation}\label{eq: stheta1}
		s_0(t ; \theta) = \mathbf{V E H B } \rho_\theta.
	\end{equation}
	
	The inner integral returning $\rho_\theta$ in (\ref{eq: ptheta}) (written explicitly in (\ref{eq: ptheta2})) is a standard parallel-beam tomographic projection of $\rho$. This integral operation can be modeled in the discretized voxel domain with a sparse projection operator akin to taking a Riemann sum\cite{sanders2015image}, which we write as $P_\theta$ so that 
	\begin{equation}\label{eq: ptheta1}
		P_\theta \rho = \rho_\theta.
	\end{equation}
	Combining (\ref{eq: stheta1}) and (\ref{eq: ptheta1}) gives rise to the 3D FFL-MPI model for the unfiltered signal at angle $\theta$ as
	\begin{equation}
		s_0(t ; \theta) = \mathbf{V E H B } P_\theta \rho.
	\end{equation}
	Observe that this sequence of operations effectively first projects $\rho$ from 3D to 2D at angle $\theta$ and then applies the 2D FFL projection model, which is diagrammed in Figure \ref{fig: 3Ddiagram}.

	Finally, we include the harmonic compression step introduced in Section \ref{sec: compress} to significantly reduce the memory requirements to solve the full 3D problem. Naturally extending the notation for the harmonic data introduced in equation (\ref{eq: bcompress}) to include an angular component gives us the compressed form of the angular data by
	\begin{equation}
		b_{DC}^\theta:= \begin{bmatrix}
			s_2^\theta \\ s_3^\theta \\ \vdots \\ s_K^\theta
		\end{bmatrix}
	\end{equation}
	in which the final model for this data is
	\begin{equation}
		b_{DC}^\theta = \mathbf A_{DC} P_\theta \rho,
	\end{equation}
	where $\mathbf A_{DC}$ is defined in (\ref{eq: DCmodel}). Assuming a set of acquired scan angles $\{\theta_1 , \theta_2 , \dots , \theta_M\}$ and the same scan sequence for each projection angle, then the full 3D model to incorporate all of the angular data into a 3D reconstruction is given by
	\begin{equation}\label{eq: modelF1}
		\begin{bmatrix}
			b_{DC}^{\theta_1}\\ b_{DC}^{\theta_2} \\ \vdots \\ b_{DC}^{\theta_M}
		\end{bmatrix}
		= \begin{bmatrix}
			\mathbf A_{DC} & 0 & \dots & 0\\
			0 & \mathbf A_{DC}& \dots & 0\\
			\vdots & \vdots & \ddots & \vdots\\
			0 & 0 & \dots & \mathbf A_{DC}
		\end{bmatrix}
		\begin{bmatrix}
			P_{\theta_1} \\ P_{\theta_2} \\ \vdots \\ P_{\theta_M}
		\end{bmatrix}\rho.
	\end{equation}

	To implement this complete operator efficiently, we need to be able to evaluate $\mathbf A_{DC}$ and each $P_{\theta_j}$ efficiently. For $\mathbf A_{DC}$, we use the same methods described in our previous work with the additional compression modification\cite{sanders2025physics}. For $P_{\theta_j}$, we use the sparse operator developed in previous works\cite{sanders2017recovering,sanders2015image}. Both of these operators are accelerated by evaluation on GPUs.

	\section{Methods}
	\subsection{Data Acquisition with Momentum Scanner}\label{sec: data} 
	Our experimental results rely on data acquisition using a commercial MPI scanner, Momentum, by Magnetic Insight, suitable for pre-clinical imaging of phantoms and small animals. The scanner has a 5.7 T/m gradient with drive field amplitudes up to 25 mT. The scanner creates an FFL along the $y$-axis with a maximum field of view (FOV) of $6\times 12$ cm$^2$ in the $xz$ plane. The scanner's system components are rotated around the $z$-axis to create angular FFL projection data modeled in Section \ref{sec: 3DFFL}. {The drive field pushes the FFL through space at 45 kHz to create the MPI signal, while a focus field and mechanical shift moves the FFL slowly through the 2D $xz$ plane to cover the full FOV. The slow shift field rasters in a zig-zag pattern, shifting quickly back and forth along the $x$-axis and slowly along the $z$-axis.}

	The system currently implements 2D projection reconstructions using the X-space methods, and full 3D imaging is accomplished on the scanner by evaluating a CT algorithm on the sequence of angular 2D projection images. Our goal is to improve upon this 3D CT method with our full 3D MPI FFL model derived in Section \ref{sec: 3DFFL}. For additional details on the scanner acquisition parameters see our previous work\cite{sanders2025physics}.

	For each imaging data set, the Momentum scanner was used to acquire the 3D FFL data at a total of 21 projection angles. The angles were acquired at equally spaced increments over the full $180^{\circ}$ range, and for each angle two scans were acquired, one corresponding to an $x$-axis drive field and the other to a $z$-axis drive field. The drive amplitudes were set at 5 mT for both axes.
	
	\subsection{Application to Imaging in Mice}\label{sec: mouse}
	All animal studies were performed in strict accordance with the NIH guidelines for the care and
	use of laboratory animals and were approved by Institutional Animal Care and Use Committee
	at Johns Hopkins University. A hemizygous MMTV-PyMT mouse that develops mammary
	tumors (B6.FVB-Tg(MMTV-PyVT) 634Mul/LellJ - Strain\#:022974) \cite{davie2007effects} was purchased from
	Jackson Labs, Bar Harbor, ME. The mouse was fed a normal diet and water ad libitum and
	maintained in the normal 12 hours of light and dark. The tumored mouse was intravenously
	injected with pegylated Synomag® (SPEG -1mgFe, lot \# 25724105-01) nanoparticles and was imaged in the MPI scanner after 72 hours under 2\% isoflurane anesthesia.

	\subsection{Numerical Methods} \label{sec: numerical}
	Our iterative model-based reconstruction is fundamentally derived from that written in equation (\ref{eq: modelF1}). The two operations on the right side of (\ref{eq: modelF1}) we denote as
	\begin{equation}
		\mathbf A := 
		\begin{bmatrix}
			\mathbf A_{DC} & 0 & \dots & 0\\
			0 & \mathbf A_{DC}& \dots & 0\\
			\vdots & \vdots & \ddots & \vdots\\
			0 & 0 & \dots & \mathbf A_{DC}
		\end{bmatrix}
		\begin{bmatrix}
			P_{\theta_1} \\ P_{\theta_2} \\ \vdots \\ P_{\theta_M}
		\end{bmatrix} ,
	\end{equation}
	and we use $\mathbf b$ to denote the data vector on the left side. Then the objective function we solve for the image reconstruction is a Tikhonov regularization model given as
	\begin{equation}\label{eq: nummodel}
		\min_x \| \mathbf A x - \mathbf b \|_2^2 + \lambda \| \mathbf T x \|_2^2,
	\end{equation}
	where $\mathbf T$ is a finite difference operator \cite{sanders2020effective}. The left term in (\ref{eq: nummodel}) is the data fidelity term, and the right term is the regularization term with scalar parameter $\lambda >0$ that reduces noise and artifacts in the solution due to the ill-posed nature of the problem \cite{kaipio2006statistical, calvetti2000tikhonov}.
	
	To solve this problem numerically, we implement a fixed step length gradient descent with a heavy ball-like acceleration \cite{su2016differential}. Then at the $k+1$st iteration, our updated solution $x^{k+1}$ is given by
	\begin{equation}\label{eq: iters}
		\begin{split}
			y^{k+1} & = x^k + \frac{k-1}{k+2} (x^k - x^{k-1})\\
			x^{k+1} & = y^{k+1} - \tau \left[\mathbf A^* (\mathbf A y^{k+1} - \mathbf b) + \lambda \mathbf T^* \mathbf T y^{k+1} \right],
		\end{split}
	\end{equation}
	where $\tau = (\| \mathbf A^* \mathbf A \|_2 + \lambda \| \mathbf T^* \mathbf T \|_2)^{-1}$. The reason for choosing the simple gradient method as opposed to say a conjugate gradient method, is that we can project the solution within each iteration to remove imaginary and negative values, which does not have a real physical interpretation.

	\section{Results}

	\subsection{Computational Time Benchmarking}
	We tested our algorithms evaluation run time for a typical 3D Momentum scan. The run time will vary depending on the scanning parameters, such as FOV, drive amplitude, and number of scan angles. For this scan, the FOV was 6 cm in each axis, the drive amplitude was 5 mT, and there were 31 equally spaced angles. The imaging mesh was set to $128^3$. The code is run in MATLAB and uses a single GPU with 16 GB of memory. We have found that 8 GB of memory to be insufficient for our implementation. The precise implementation and run time of all of the 2D operators was reported in our previous work \cite{sanders2025physics}.
	
	\begin{table}[ht]
		\centering
		\begin{tabular}{|c|c c|}
			\hline
			\textbf{Operator} &	forward & adjoint \\ \hline
			$P_\theta$  (all angles simultaneous)  & 0.0419 & 0.0077 \\
			$\mathbf A_{DC}$ (one angle)    & 0.0132 & 0.0539  \\
			$\mathbf A_{DC}$ (31 angles)    & 0.4095 & 1.6724  \\ \hline
			total time for single iteration	  &  2.0973 & \\
			total time for 150 iterations  & 314.6016 & \\ \hline
		\end{tabular}
		\caption{\label{table: time2}Average run time (in seconds) of each of the operators evaluated in MATLAB over 20 trials on a $128^3$ test image.}
	\end{table}

	The run times of each of the operators involved are reported in Table \ref{table: time2}. The $\mathbf A_{DC}$ operator is reported for both the run time for single application of the operator and for all 31 angles, since each angle is evaluated sequentially by design. This reported time includes the total time for both the $x$ and $z$ collinear scans. The projection operators for all angles are evaluated simultaneously as a single operation, which is a sparse GPU matrix \cite{sanders2015image}. For a model-based iterative reconstruction, a single iteration requires one application of the matrix-based operation and one application of its adjoint (see Section \ref{sec: numerical}). When this is all added together for our problem, we see in the table that a single iteration would require roughly 2 seconds, which translates to just over 5 minutes for 150 iterations. However, this will slightly underestimate the true evaluation time, because our model-based iterative reconstruction also includes a regularization term in (\ref{eq: nummodel}). 

	\begin{table}[ht]
		\centering
		\begin{tabular}{|c|c | c| c |}
			\hline
			\textbf{FOV} &	\textbf{Drive Amp.} & \textbf{\# Angles}& \textbf{run time} \\ \hline
			
			$6\times 6\times 6 \, cm^3$ & 23 mT & 35 & 182.3 s\\
			$6\times 6\times 6 \, cm^3$ & 5 mT & 21 & 530.7 s\\
			$6\times 6\times 12 \, cm^3$ & 5 mT & 21 & 1665.1 s\\ \hline
		\end{tabular}
		\caption{\label{table: speed}Reconstruction times in seconds observed for 150 iterations of our model-based algorithm on 3 different data sets with unique scanning parameters.}
	\end{table}

	Table \ref{table: speed} reports the actual time observed to complete 150 iterations of our algorithm for 3 distinct data sets, each with unique scanning parameters. We can see the shortest run time is roughly 3 minutes, while the longest is roughly 28 minutes. Hence, the evaluation time is highly dependent on the scan parameters, while nevertheless still capable of reconstructing in reasonable time frames for practical use.

	\subsection{Sensitivity Data and Detection Limit}

\begin{figure}[ht]
		\centering
		\includegraphics[width=0.7\textwidth]{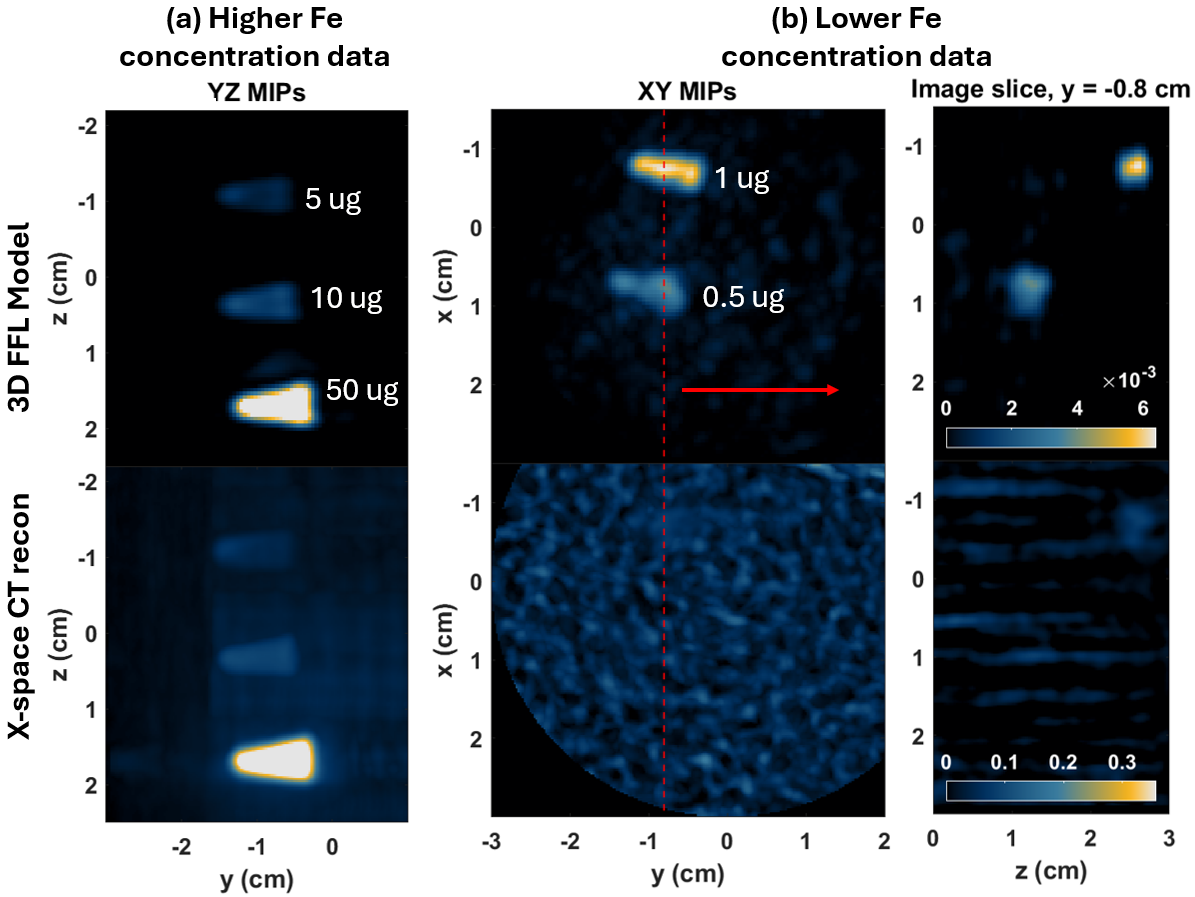}
		\caption{Results on sensitivity series data comparing the 3D FFL model (top row) with the X-space CT projection reconstruction (bottom row). The images on the left show the results from the higher concentration samples (5-50 ug) and the right images show the results from the lower concentration data (0-1 ug). The left two columns show maximum intensity projections (MIPs), and the right images show a single slice of the 3D volume, where the location is indicated by the red dashed line in the middle column.}
		\label{fig: cal}
	\end{figure}

	A series of samples were prepared in water at a volume of 167 uL in plastic Eppendorf tubes. Each sample was mixed with different quantities of Synomag PEG MNPs ranging up to 50 ug.  The data was split into two separate batches for the scans. One scan was used to image the higher Fe concentration samples at 5, 10, and 50 ug of Fe, and the second scan was used for the lower Fe concentration samples at 0 (control), 0.5, and 1 ug. The Momentum scanner was used to acquire the 3D MPI data as described in Section \ref{sec: data}. Shown in Figure \ref{fig: cal} are the resulting image reconstructions comparing the new 3D FFL model with the X-space CT reconstructions.
	
	For the higher concentration data set (left images) all MNP samples are clearly imaged in both reconstructions. However, only the model-based reconstruction is able to resolve the two MNP samples in the lower concentration data set (middle and right images). The cross-sectional images shown on the far right indicate that the 1 ug sample is partially visible in the X-space CT reconstruction, although it is practically level with the background noise. Based on these quick observations, we could argue that the 3D FFL model provides a roughly 10x improvement in the Fe detection limit.

	\begin{figure}
		\centering
		\includegraphics[width=0.75\textwidth]{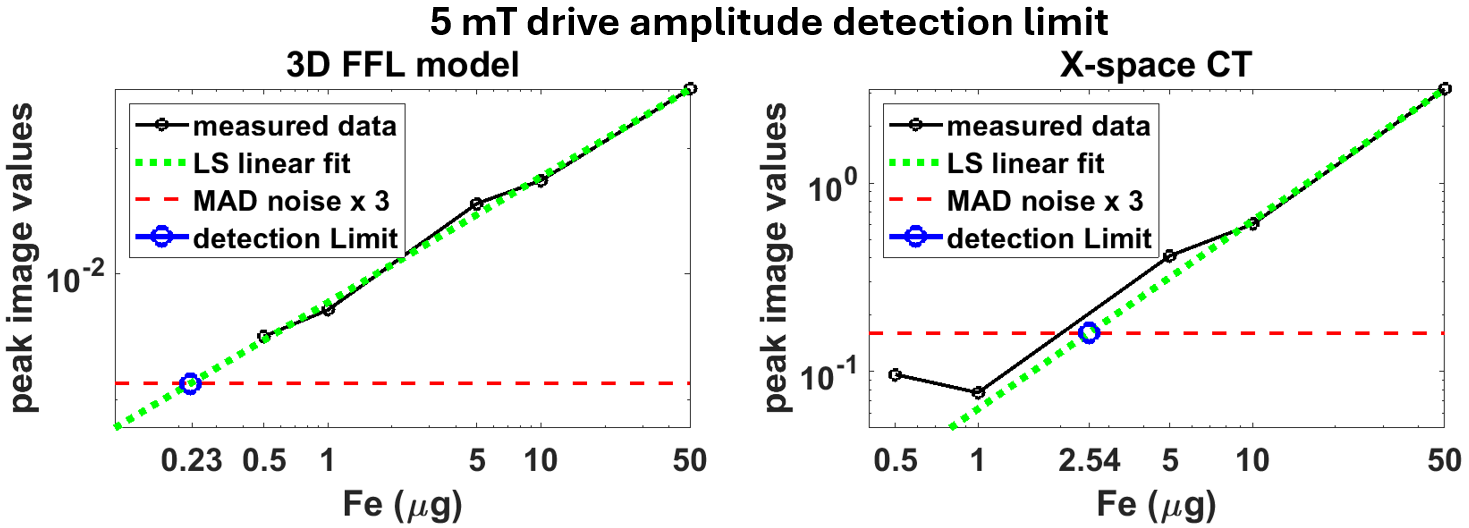}
		\caption{Analysis of the detection limit with the new model vs. the X-space CT reconstruction using the sensitivity data sets. The detection limit for the new model was determined to be 0.23 ug, while for the X-space CT reconstruction the limit was 2.54 ug. This corresponds to an 11.2x improvement in the detection limit with the new model. Note that a higher drive amplitude of 25 mT would theoretically improve both detection limits by another factor of 25 (see footnote).}
		\label{fig: det}
	\end{figure}

	A thorough analysis of the detection limit is presented in Figure \ref{fig: det}, where we estimate the detection limit for each method by estimating where the background noise level and the MNP image values intersect. Using the measured image values at the higher concentration data, a least squares linear curve (green plot) was fit to the image values as a function of the Fe content. The background noise was measured using 5 void rectangular regions in the images away from the MNP samples. The final noise estimate was taken to be the mean absolute deviation (MAD) of these regions. We defined the detection threshold to be 3 times the mean noise level, which is statistically significant\footnote{We do not assume any particular distribution for the noise, which is likely colored in both reconstructions due to the MPI physics and CT inversion.}. Therefore the detection limit was taken to be the point where the linear fit curve crossed the 3x noise level (blue circle). We can see for the 3D FFL model this is at 0.23 ug and at 2.54 ug for the X-space CT reconstruction, which corresponds to a 11.2x improvement in the detection limit with the new model\footnote{Note that the 5 mT excitation is 1/5th the full power of the scanner, and the values here do not represent the sensitivity limit of the scanner. The sensitivity limit at full power would improve by 25x as deduced from Theorem 1 in \cite{sanders2025multi}.}.

	\subsection{Intravenous MNP Mouse Imaging with High Dynamic Range}\label{sec: mouseR}

	\begin{figure}
		\centering
		\includegraphics[width=0.85\textwidth]{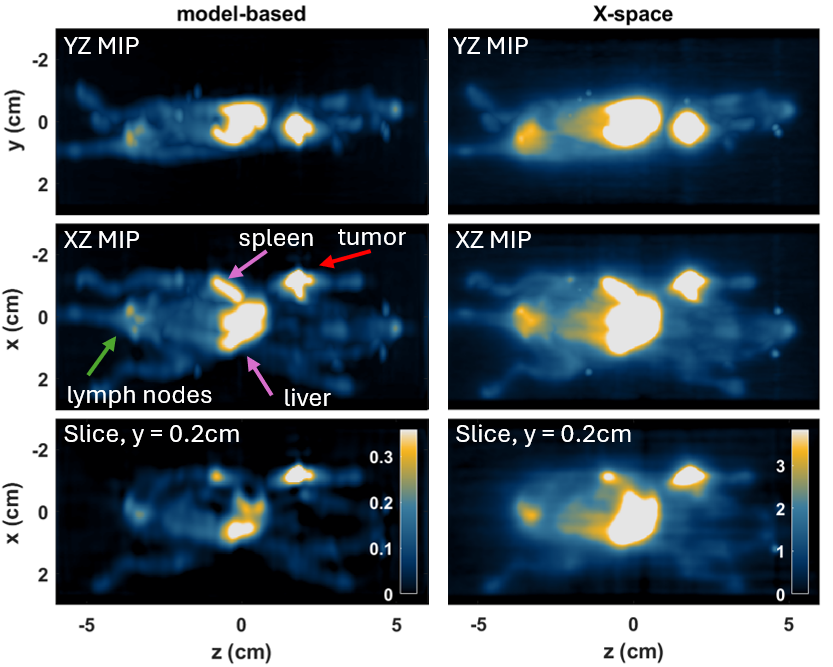}
		\caption{3D image reconstruction results on a mouse that was intravenously injected with MNPs using the new model (left) and the 3D CT reconstruction (right) from the X-space projection images. The top two rows show maximum intensity projection images (MIPs), which are formed by taking the maximum value along the respective axes ($y$-axis in the top row and $x$-axis in the middle row). The bottom row shows a single cross-sectional slice at $x=0.2$ cm. For each image the contrast is adjusted to show the top limit at 0.4 times the maximum.}
		\label{fig: mouse1}
	\end{figure}

	
	A live mouse was intravenously injected with 1 mg of MNPs and a 3D data acquisition was performed in the MPI scanner after 72 hours, as outlined in the methods section \ref{sec: mouse}. The infused MNPs circulate and can accumulate in macrophages in the liver, spleen, lymph nodes, and bone marrow, as well as inflammatory cells in the tumor microenvironment \cite{daldrup2011mri}. This experiment evaluates the reconstruction sensitivity and ability to recover weak signals adjacent to strong background compartments.
	
	The 3D reconstructed images of this mouse are shown in Figure \ref{fig: mouse1}, which compares the model-based reconstructions (left) with the CT reconstruction (right) from the sharpened X-space projection images \cite{lu2015linearity}. All images are contrast adjusted so that the maximum display is set at 0.4 times the maximum of the total volume due to the large dynamic range (the color bar in the bottom images is valid for each column). In addition, 3D volume renderings of the reconstructions are shown in Figure \ref{fig: mouse3D}.

	The model-based reconstruction yields notably better resolution and less background haze caused by large signals in the liver and spleen (indicated by the purple arrows). Due to significant accumulation of MNPs here, the resulting bright region leaves an unwanted haze over the entire X-space CT image, which is very apparent in the 3D volume rendering. This creates a challenging scenario in MPI of high dynamic range \cite{fernando2024focused}, where the high background signal from the liver can wash out neighboring regions of potential interest with low signal, such as the lungs. The model reconstruction significantly reduces this signal haze such that the liver signal is confined solely to the liver region, furthermore, each of the individual liver lobes can be visualized. Using the model reconstruction, each of the air-filled lungs can clearly be visualized as dark regions in vivo despite being directly adjacent to the bright liver. The lungs were confirmed to have minimal MNP accumulation (293x lower) compared to the combined liver/spleen signal with ex vivo MPI. 

	
	The model reconstruction additionally improves resolution in other regions, such as near the base of the tail, where 3 sciatic lymph nodes are clearly distinguished (indicated by the green arrow), that are not resolved with the original approach. A second bright region is detected in the upper arm of the mouse (red arrow), which is a known palpable mammary tumor that also significantly accumulated MNPs. In the sequentially processed CT reconstruction,
	the high intensity signal from the tumor creates a haze that extends beyond the tumor bounds, which can introduce uncertainty when performing quantification. Our new model reconstruction significantly improves the resolution and further defines the MPI signal within the tumor boundary.


	\section{Discussion}
	This work introduces a full 3D field-free line (FFL) image reconstruction model for magnetic particle imaging (MPI), representing a significant extension of previously developed 2D projection models. By combining a physics-based MPI signal model with tomographic projection operators and harmonic domain compression, our method enables direct reconstruction from time-domain data across multiple projection angles, all within a computationally feasible framework.
	
	
	Experimental comparisons with conventional 3D X-space CT reconstructions underscore the advantages of our approach. Phantom studies demonstrate over 11x improvement in Fe detection sensitivity, while in vivo imaging in mice reveals enhanced resolution and significantly reduced background haze, especially in high dynamic range scenarios involving bright liver signals adjacent to low-intensity regions like the lungs. This reduction in signal bleed through improves visualization of anatomical structures and may facilitate more accurate quantification of signal localized to small tumors or lymph nodes.
	
	While our method shows substantial improvements, some limitations could be addressed in future work. While the Langevin-based MNP response model is widely accepted, future work may further refine accuracy by incorporating relaxation effects or inter-particle interactions, though at additional computational cost. Additionally, while reconstructions were demonstrated on preclinical data, scaling this method to clinical field-of-view sizes or integrating it into real-time imaging pipelines will require further development.
	
	\section{Conclusions}
	We have developed and validated a new 3D image reconstruction method for MPI based on a field-free line (FFL) model combined with tomographic and harmonic compression techniques. This framework enables direct volumetric reconstruction from raw time-domain MPI data without relying on intermediate 2D projections or computationally intensive system matrix calibration.
	
	Our method significantly improves spatial resolution and detection sensitivity while reducing background artifacts in challenging high dynamic range imaging scenarios. The framework is computationally efficient, runs on standard GPU hardware in 5 to 10 minutes, and offers a flexible foundation for future extensions such as enhanced MNP modeling.
	
	These advances represent a step forward in making high-quality 3D MPI practical for preclinical and potentially clinical applications, with the potential to improve diagnostic accuracy across a variety of biomedical imaging tasks.

	\appendix
	\section*{Appendix}
	\section{Analysis of the Digital Compression (DC)}\label{sec: companaly}

	For our received signal $\vec s \in \R^N$ and an arbitrary frequency $\omega\in \{ 0, 1, \dots, N-1\}$, the (unitary) Fourier transform of the received signal at this frequency is given by
	\begin{equation}
		\hat s_\omega = (\F \vec s)_\omega = \tfrac{1}{\sqrt{N}}\sum_{j=0}^{N-1} s_j e^{i2\pi \omega j/N}.
	\end{equation}
	Since this transform is unitary (and by Parceval's theorem), we have
	\begin{equation}
		\sum_{\omega = 0}^{N-1} |\hat s_\omega |^2 = \| \vec s \|_2^2 . 
	\end{equation}
	Therefore, it is apparent how to estimate the amount of energy (information) in the signal that is lost in the compression step: it is directly proportional to the squared sum of the coefficients that are dropped in the compression step. Let us define $\mathbb S \subset \{ 0, 1, \dots, N-1\}$ to be the set of frequency coefficients that are kept in the compression step, in which case our data vector $b_{DC}$ in (\ref{eq: bcompress}) could be alternatively written as 
	\begin{equation}
		b_{DC} = \{ \hat s_\omega \}_{\omega \in \mathbb S} .
	\end{equation}
	Then we define the information retained in the compression as 
	\begin{equation}
		I(\vec s, \mathbb S) := \frac{\sqrt{\sum_{\omega \in \mathbb S} |\hat s_\omega|^2 }}{ \| \vec s \|} .
	\end{equation}
	\begin{figure}
		\centering
		\includegraphics[width=0.7\textwidth]{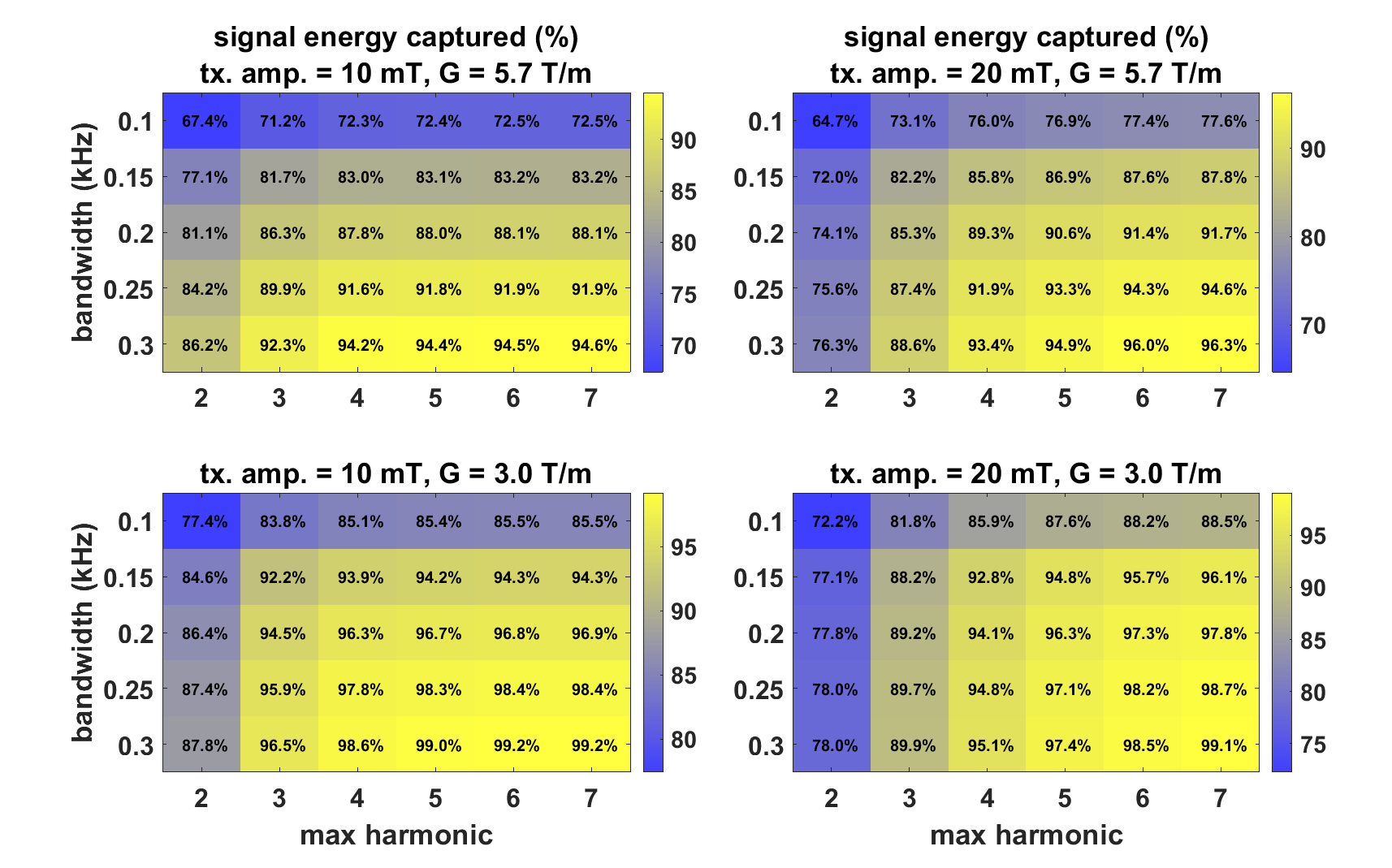}
		\caption{Signal energy (information) captured in the harmonic Fourier compression as a function of harmonics and bandwidth, while also showing dependence in the separate plots on the transmit amplitude and magnetic field gradient strength.}
		\label{fig: DC}
	\end{figure}
	
	Figure \ref{fig: DC} shows the amount of information retained in a simulated MPI signal from a typical Momentum scan. The information is shown in these image plots as both a function of the bandwidth retained around each harmonic and the number of harmonics retained. There are 4 distinct plots in the Figure, each pertaining to a unique set of scanning parameters. Observe that using up to 0.3 kHz of bandwidth up to harmonic number 5 already retains roughly over 95\% of the information in all cases. Notice also that at higher drive amplitudes the compression is less efficient, which can be deduced from the harmonic analysis theory presented in \cite{sanders2025multi}.

	\section{Notation}\label{sec: notation}
	Key mathematical notations used are provided below for convenience. 
	
	\begin{itemize}\itemsep -.2cm
		\item $\vec x = (x,y,z)$ is the spatial location variable.
		\item $t$ is the time variable.
		\item $\rho(\vec x)$ is the MNP density map.
		\item $\vec b_1(\vec x)$ is the receive coil sensitivity map.
		\item $\mathcal L(x) = \coth (x) - \tfrac{1}{x}$ is the Langevin function.
		\item $H(\vec x, t)$ is the time varying magnetic field.
		\item $m$ is the magnetic moment of the MNPs.
		\item $\beta = \frac{\mu_0 m}{\kappa_B T}$ is the conglomerate MNP particle constant, where $\kappa_B$ is the Boltzmann constant, $T$ is temperature, and $\mu_0$ is the vacuum permeability. 
		\item $s_0(t)$ is the unfiltered MPI received signal.
		\item $s(t)$ is the receive chain filtered MPI received signal. 
		\item $\vec \xi(t)$ is the FFR, i.e. $H(\vec \xi(t), t) = 0$. 
		\item $\vvec{h} (\vec x)$ is the $3\times 3$ MPI tensor PSF given by
		\begin{equation}\label{eq: PSF}
			\begin{split}
				&\vec{\vec{h}} (\x) := 
				\Bigg[ \mL'(\| G \x \|/H_{sat}) \frac{G \x \x\T G\T }{\| G \x \| H_{sat}} +   \mL (\| G \x \|/H_{sat}) \left( I - \frac{G \x \x\T G\T }{\| G \x \|^2}
				\right) \Bigg] \frac{G}{\| G \x \|}.
			\end{split}
		\end{equation}
		\item $\mathbf A_{DC}$ is the data compressed MPI model operator. 
		\item $P_\theta$ is the parallel beam projection operator at angle $\theta$. 
	\end{itemize}


\end{document}